# The Third State of the Schelling Model of Residential Dynamics

**Itzhak Benenson[1], Erez Hatna[2]**

[1]Department of Geography and Human Environment,

Tel Aviv University, Tel Aviv, Israel

[2]Department of Environmental Sciences,

Wageningen University, Wageningen, The Netherlands

bennya@post.tau.ac.il, erez.hatna@wur.nl

**Abstract**

The Schelling model of segregation between two groups of residential agents (Schelling 1971; Schelling 1978) reflects the most abstract view of the non-economic forces of residential migrations: be close to people of "your own". The model assumes that the residential agent, located in the neighborhood where the fraction of "friends" is less than a predefined threshold value **F**, tries to relocate to a neighborhood for which this fraction is above **F**. It is well known that for the equal groups, depending on **F**, Schelling's residential pattern converges either to complete integration (random pattern) or segregation.

We investigate Schelling model pattern dynamics as dependent on **F**, the ratio of the group numbers and the size of the neighborhood and demonstrate that the traditional integrate-segregate dichotomy is incomplete. In case of unequal groups, there exists the wide interval of the **F**-values that entails the third persistent residential pattern, in which part of the majority population segregates, while the rest remains integrated with the minority. We also demonstrate that Schelling model dynamics essentially depends on the description of agents' residential behavior. To obtain sociologically meaningful results, the agents should be satisficers, and the fraction of the agents who relocate irrespective of the neighborhood state should be non-zero.

# 1. Introduction

## 1.1. The basic framework of the Schelling model

The Schelling model of segregation (Schelling 1971; Schelling 1978) is the long-standing engine of agent-based simulations in the social sciences. In the urban context it reflects the very basic and most abstract view of the non-economic forces of residential migrations: be close to people of "your own".

Generally, the model follows the "pull-push" or "stress-resistance" hypothesis of residential migration formulated several years earlier (Brown and Moore 1970; Speare 1974). Formally, "residential agents" occupy cells of rectangular "residential space", not more than one agent in a cell. Agents belong to one of two types, which we denote below as **B**(lue) and **G**(reen). Agent "knows" the state of every cell in a neighborhood around their location – whether the cell is occupied and, if yes, the type of the agent in it.

The model's basic assumption is as follows:

*An agent, located in the center of a neighborhood where the fraction of friends **f** is less than a predefined threshold value **F**, i.e. **f** < **F**, wants to relocate to a neighborhood for which **f** ≥ **F** (Schelling 1978), p. 148)[1].*

Thomas Schelling employed the model to examine "some of the individual incentives and individual perceptions of difference that can lead collectively to segregation" (Schelling 1978), p. 138). To do so he placed equal numbers of dimes and pennies on squares of a chessboard-size sheet of paper and assumed that every dime or penny wanted "something more than one-third" of his neighbors within the 3x3 neighborhood to be like himself. Any dime or penny whose neighborhood did not meet this condition got up and moved to the nearest square that satisfied his demand (Schelling 1978), pp. 147-148).

---

[1] The number of neighbors in the neighborhood is finite, so it does not matter whether "less" or "less or equal" is employed in the definition. Different papers use different versions of the inequality, and we should keep this in mind when talking about the Schelling model results.

As it is commonly known that regardless of the initial pattern and the order of moves, the dimes and pennies in Schelling's experiment quickly segregate and the pattern stalls. In this and similar experiments Schelling demonstrated (and usually inspired the reader to do likewise) that for **F** above one third the patterns in his model converge to segregate equilibrium. He also mentions that the properties of the equilibrium pattern depend on the value of **F**, and that **F** can differ for dimes and pennies (Schelling 1978), p. 153-154).

### 1.2. The state-of-the-art of the Schelling model studies

#### 1.2.1. The standard settings and basic dichotomy of patterns

The settings established by Thomas Schelling (Schelling 1978) are as follows:

S1.   The city is a grid of cells and each cell's neighborhood is 3x3 square around it, truncated by the city boundaries if close to them.

S2.   The city is populated by agents, each belonging to one of two groups of equal size. We denote these groups as **B**(lue) and **G**(reen), and, thus, **B:G** = 1:1.

S3.   Initial distribution of agents on the grid is random, and the fraction of vacant cells is sufficient for relocations.

S4.   Model dynamics are considered in discrete time. At every iteration every agent estimates the fraction **f** of the neighbors of its type (friends) within the neighborhood (empty places are ignored); agents are considered in random order.

S5.   There exists a threshold value of **F** of the fraction of friends within the neighborhood, which is common for all agents. An agent located at a center of a neighborhood where **f** < **F** decides to leave the cell and relocate to the *closest* empty cell satisfying the condition **f** ≥ **F**.

S6. Information about a cell left by an agent becomes immediately available to all agents.

In what follows we call S1 – S6 the "standard settings" of the Schelling model.

Note that the standard settings are incomplete, for example, they are insufficient to resolve the situation when none of the vacancies satisfies the condition **f** ≥ **F**. As a result, one should be careful when comparing results of different studies.

The most general view of the Schelling model dynamics is as follows:

- In case of weak tendency to reside in a friendly neighborhood (low **F**), the model's residential pattern converges in time to the pattern that is visually indistinguishable from the random one (Figure 1a).

- In case of strong tendency to reside in a friendly neighborhood (high **F**), the model's residential pattern converges in time to the segregate one, characterized by one or more homogeneous patches of **B**- and **G**-agents separated by unpopulated boundaries (Figure 1b)

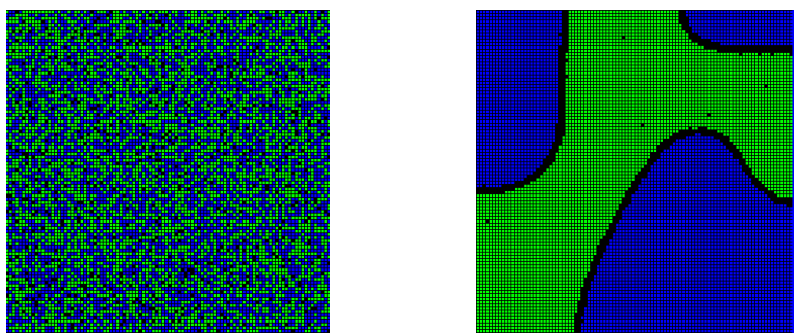

(a) (b)

Figure 1: Persistent residential patterns of the Schelling model in case of 95% density of agents: (a) **F** = 0.2 – the residential pattern cannot be distinguished from the random one; (b) **F** = 0.8 – the residential pattern is segregate.

### 1.2.2. Fine and rough effects in the Schelling model

The Schelling model is formulated "verbally", and is studied mainly by simulation[2]. Several studies implement the model with minor deviations from the standard settings. Flache and Hegselmann (Flache and Hegselmann 2001) weakened condition S1 and investigated the model for irregular partition of the plane into polygonal units and definition of neighborhood on the basis of polygons' adjacency. They demonstrate that convergence to the segregate or the random pattern is characteristic of the model irrespective of size and form of the neighborhood. Laurie and Jaggi (Laurie and Jaggi 2002; Laurie and Jaggi 2003) weaken the condition S1 in another way and investigated the Schelling model dynamics in respect of the size **R** of the cell's neighborhood. They claim that for given **F**, the number of homogeneous patches in the patterns decreases, and they become larger with the growth of **R**. Portugali et al (Portugali, Benenson et al. 1994; Portugali and Benenson 1995) weakened condition S5 and essentially extended Shelling's aforementioned remark regarding group-specific **F** (Schelling 1978, p. 153-154) by demonstrating that symmetric behavior of the **B**- and **G**-agents behavior is not at all necessary for segregation. They show that the model's residential pattern can converge to segregate even if the agents of one type only react to neighborhood structure, the agents of another type being fully tolerant of it.

Note that the Schelling model is not structurally stable and its dynamics can qualitatively change when one of the conditions S1 – S6 is altered. The Netlogo version of the model, for example (http://ccl.northwestern.edu/netlogo/), displays *unstable* spatial patterns for the values of **F** close to unit. However, as one can easily find by checking the model code - the agents in the Netlogo implementation do not verify the condition **f ≥ F** when choosing the vacant cell and relocate to the arbitrary vacancy.

---

[2]R. Koenig's application http://www.entwurfsforschung.de/RaumProzesse/Segregation.htm can be a starting point. More complex situations involving agents differing in several characteristics can be investigated with the help of M. Fosset's application http://sociweb.tamu.edu/vlabresi/vlab.htm

Several papers extend the Schelling model to a multi-dimensional characterization of agents, that is, they vary condition S2. Some of the papers (Benenson 1999; Portugali 2000; Fossett 2006b; Benard and Willer 2007) vary condition S2 by consider agents who, in addition to the Boolean "color", are characterized by the continuous "economic status", which can be compared to the cells' price. As might be expected, the system dynamics in this case becomes more variable, and depends on way the agents' reactions to the neighbors' color and status are combined in utility function.

Preserving Schelling's Boolean view of the agents' properties, Benenson (Benenson 1998) characterizes agents by K > 1 Boolean characteristics, which results in $2^K$ instead of just 2 "phenotype" groups. Given the size of the grid, these groups become too many with the growth of K, and the model dynamics exhibits unstable but persistent regimes in which a few homogeneous groups of a given phenotype repeatedly segregate, persist for some time, and then dissolve and are substituted by the segregate groups of the other phenotypes.

A qualitatively different way of varying the S5 condition is to assume that unsatisfied agents exchange their places rather than search for a vacancy over all unoccupied cells. This view has recently attracted the attention of several researchers (Pollicott and Weiss 2001; Zhang 2004), who also assume that the preferences of agents are non-monotonous, and they might prefer neighborhoods with a low fraction of foreigners to those occupied exclusively by friends. The dynamics of the model patterns in this case is more complex than the random–segregate dichotomy. However, this model qualitatively differs from the standard settings in its mathematical properties and we thus consider this formulation as different from Schelling's original.

Another , "Bounded Neighborhood" version of the Schelling model (Schelling 1978, p. 155), is based on partition of grid space into blocks larger than a cell. No matter where the agent is located in the block, it reacts to the fraction of friends all over it, and leaves the block if the fraction of friends there is below **F**. Different agents, however, react to different threshold fractions of friends, and the main parameter of the "Bounded Neighborhood" model is distribution of agents' **F**-values.

The analytical (and non-spatial) investigation of this version of a model was started by Schelling (Schelling 1978) and continued by Clark (Clark 1991; Clark 1993; Clark 2002; Clark 2006), who demonstrated that similar to the basic version of a model, the block population remains mixed if the fraction of the agents able to stay in a block despite a low fraction of friends is sufficiently high.

A spatial version of the "Bounded Neighborhood" model, in which agents migrate among cells of the blocks, has been investigated in depth in a series of recent simulations by Fosset (Fossett and Waren 2005; Fossett 2006a; Fossett 2006b), who also directly relates the model's results to residential patterns in the real-world cities. In terms of conditions S1-S7, Fosset varies condition S2 and considers the population of agents of the three residential groups – White, Black and Hispanic, who, in addition, differ in economic status. The aggregate units are 7x7 cell blocks and the city is represented by the 12x12 grid of blocks. Fosset investigates the relationship between ethnic and status segregation, and concludes that this interaction can limit our ability to integrate the ethnic groups. Specifically, he points out that reductions in housing discrimination may not necessarily lead to large declines in ethnic segregation.

Several attempts to apply the Schelling model to the real-world situation could not avoid varying several of the conditions S1 – S6 simultaneously (Benenson, Omer et al. 2002; Koehler and Skvoretz 2002; Bruch and Mare 2006). They all provide likelihood approximations of the segregate or more complex residential distributions observed in the cities. In what follows we limit ourselves to abstract models and do not go into greater detail of these implementations.

To conclude, some deviations from the standard settings do not influence the basic random-segregate dichotomy, while some alter it. We thus need some general insight on the variety of the patterns in the Schelling model. Recently, the first step toward this view was suggested by Vinkovic and Kirman (Vinkovic and Kirman 2006), who define a continuous analog of the Schelling model. Namely, they represent the eight outer cells of the 3x3 neighborhood as eight equal sectors, each reflecting the state of the agent in the corresponding cell. Then they "weaken" the discrete representation of space by cells and consider this angle a continuous characteristic, thus enabling the description of the dynamics of the borders of clusters of agents of both types by means of differential equations.

The analysis of the continuous analog of Schelling model clarifies two basically different groups of relocation rules: one specifying Schelling's system as representing "solid" matter, the other specifying it as representing "liquid" matter. The rules that allow relocation to the *better location only* result in solid-like dynamics – the system stalls when converging to the pattern in which none of the agents can improve their state. The rules that allow relocation to the *cells of the same utility* result in liquid-like dynamics, when the system pattern can change following relocation of the agents who do not improve their state. The numeric examples, presented by Vinkovic and Kirman (2006) demonstrate that the solid–liquid dichotomy of the agents' relocation rules can be also valid for the discrete version of the Schelling model.

This paper is a first step towards systematic investigation of the qualitative properties of the traditional - discrete - version of the Schelling model. We study model dynamics as dependent on (1) the threshold fraction of friends **F**; (2) varying size of neighborhood **r** and (3) varying ratio of **B** and **G** population groups, and reveal additional, not-random and not-segregate persistent model pattern. The solid-liquid dichotomy of the relocation rules is relevant for the discrete model just as it is for the continuous one; moreover, we demonstrate that only the rules that entail liquid-like dynamics have meaningful real-life interpretation.

## 2. Detailed description of the model

In what follows we consider the Schelling model in discrete time and space, where a *city* is a *torus* grid of cells, populated by **B-** and **G-**agents, who act in discrete time **t** = 0, 1, 2, … (we omit index **t** unless the meaning becomes ambiguous). At each time step (below, iteration), every agent decides whether and where to relocate. Agents are considered in a random order, which is established anew with each iteration. We follow Schelling's condition S6, and assume that an agent observes the changes of system *immediately after* they happen. Formally, this view corresponds to asynchronous updating (Cornforth, Green et al. 2005).

### 2.1. Model rules and definitions

Let us consider the torus city of cells of **NxN** size, each cell can be occupied by one residential agents and the fraction of occupied cells in the city is **d**. Agents can belong to one of two type – **B**(lue) or **G**(reen). An agent **a** located at a cell **h**, remains there, or relocates, in respect to the fraction of friends (agents of the same type) within the neighborhood of **h**. In what follows we consider as a neighborhood a square of **(2*r+1)*(2*r+1)** size with **h** in a center, excluding **h**, and call **r** the radius of the neighborhood

For an agent **a** located at cell **h** let us denote

- neighborhood of **h**, excluding **h** itself, as **U(h).**
- fraction of **a**-type *friends* among the agents located within **U(h)** as **$f_a$(h).**
- minimal fraction of friends necessary for **a**'s staying within **U(h)** as **F.**

We assume that

- An agent **a** can relocate by the reasons independent of the number of friends within the neighborhood (we call this below "random reasons") with probability **m**.

- When relocating, an agent is able to consider and compare not more than **w** vacant places.

- The awareness of the agent **a** located at **h** about the vacancy **v** does not depend on the distance between **h** and **v**.

To decide whether to relocate or not an agent **a** according to the following rules:

Step 1: *Decide whether to try to relocate*:

- Estimate **$f_a$(h)** by dividing the number of agents of **a**-type within **U(h)** by the overall number of agents there i.e., ignoring **h** and empty places within **U(h)**.
- Generate a random number **p,** uniformly distributed on (0, 1).
- If **$f_a$(h) < F** or **p < m** then memorize **$f_a$(h)** and decide to relocate, otherwise decide to stay at **h**.

Step 2: *If the decision is "to try" then search for the new location and decide whether to move there:*

- Construct set **V(a)** of opportunities by randomly selecting **w** vacancies from all vacancies existing at a moment.

- Estimate utility $u_a(v)$ of each $v \in V(a)$ as $\min(f_a(v), F)$ i.e., assuming that all vacancies with $F$ and higher fraction of friends have the same utility $F$.

- Select the vacancy $v_{best} \in V(a)$, whose utility $u_a(v_{best})$ is the highest among all $v \in V(a)$. If there are several best vacancies, choose one of them randomly.

- Move to $v_{best}$ if

    - the fraction of friends at a current location is low - $u_a(h) < F$ - and the migration would improve it - $u_a(v_{best}) > u_a(h)$

    - the reason of migration is other than unfriendly neighborhood - $u_a(h) \geq F$ - and a new neighborhood is friendly enough - $u_a(v_{best}) \geq u_a(h)$.

- Otherwise stay at $h$.

The sequence of model events is presented in Figure 2.

Let us emphasize that following Schelling's (Schelling 1978) view, model agents are "satisficers" (Simon 1982) i.e., they can migrate *between two vacancies* as far as both have a fraction of friends equal to $F$ or above. We thus might expect the liquid-like system dynamics (Vinkovic and Kirman 2006). Below we discuss the "maximizer's" view, when a utility $u_a(v)$ of the location $v$ for $a$ equals to the number of friends within $U(v)$.

Let us specify the parameters' space and the way we describe the model's results.

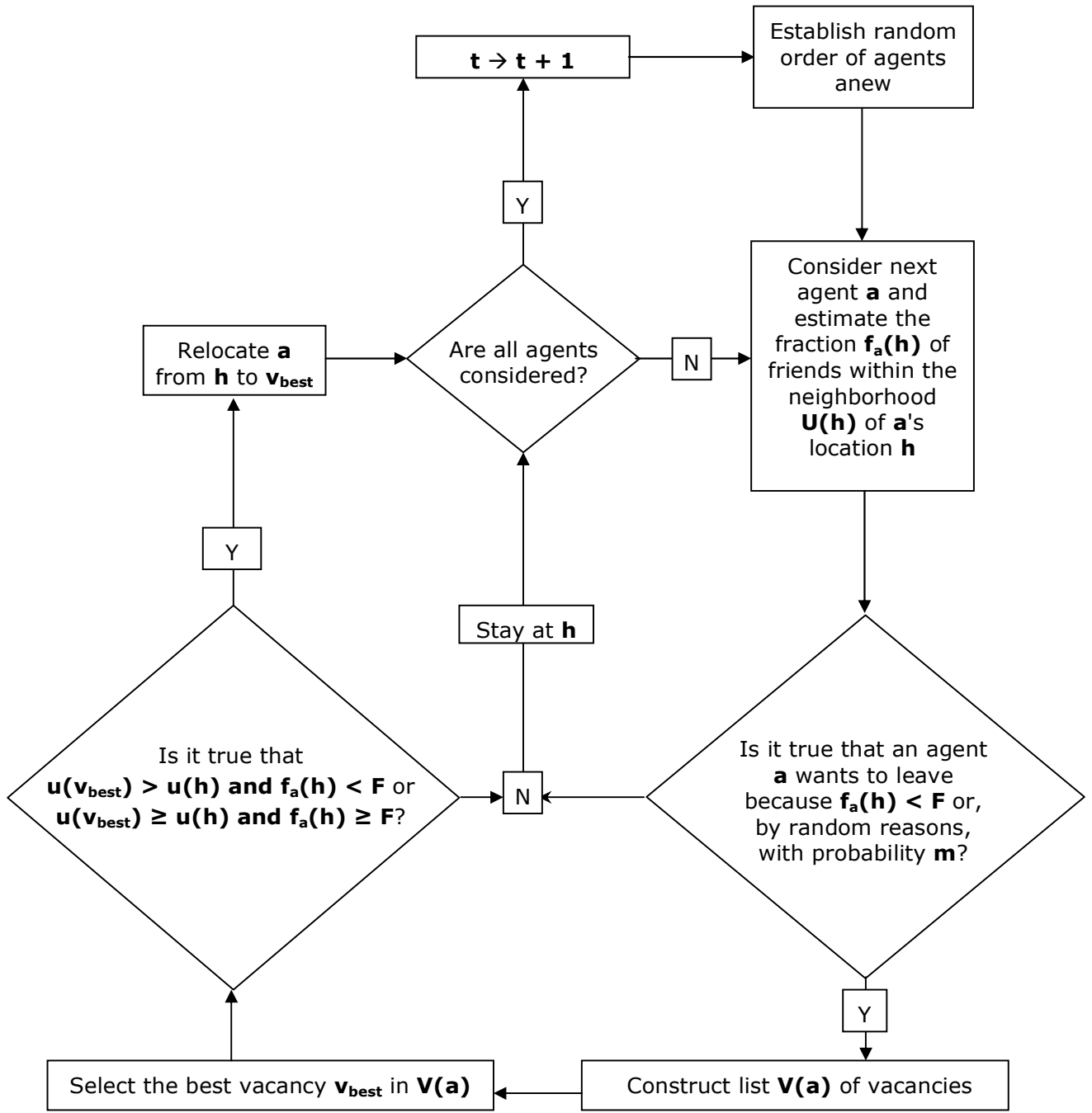

Figure 2: Flow chart of the model

## 2.2. Model settings, investigated parameters and initial conditions

The model settings are presented in Table 1:

| Table 1: The values of model parameters | |
|---|---|
| Parameter | Value and its qualitative meaning |
| Form and size of the city | **N**x**N** torus, **N** = 50 or **N** = 100 |
| Form and size of the **U(h)** | Moore neighborhood of the size **n(r) = (2*r+1)*(2*r+1) − 1**, **r** = 1 to 4 in section 3, **r** = 1 to 3 in section 4 |
| Rate **m** of spontaneous leave per iteration | **m** = 0.01 |
| Initial density of agents | **d** = 0.98 (very high) |
| Maximum number of vacancies **w** considered by an agent for relocating | **w** = 30 (sufficiently high) |
| Fraction of minority **β = B:(B+G)** | β ≤ 0.5 - either group numbers are equal or the **B-**agents are the minority |
| Initial conditions | **Random** and **Segregate** |

We investigate the model's dynamics for the following set of parameters:

- Moore neighborhood of the radius **r** = 1 to 4.

- The fraction **β** of the minority **B**-group as varying on [0, 0.5].

- The threshold intolerance **F** as varying on [0, 1].

Note that for a neighborhood of a radius **r**, the number of friends can vary between **0** and **n(r) = (2*r+1)*(2*r+1) − 1**. That is, for given **r** we can study the system's dynamics for any series **F$_k$** of **n(r) + 1** values of **F**, satisfying the condition **(k − 1)/n(r) < F$_k$ ≤ k/n(r)** (**k = 0, 1, …, n(r)**). To avoid the problems of rounding, in what follows we employ the series

**F$_k$ = k/n(r) - 1/(2*n(r))**, where **k = 0, 1, …, n(r)**     (1)

**F = F$_0$** provides the same results as **F = 0**, and corresponds to the case when agents are absolutely insensitive to the neighbors, **F = F$_1$** provides the same results as **F = 1/n(r)** and is characteristic of the agents who demand one friend at least in the fully occupied neighborhood, etc.

The initial random pattern of agents (as in Figure 1a) is established by allocating agents in exactly **Round(d*N*N)** cells. **Round(β*d*N*N)** of agents are randomly set of type **B** and the rest **Round(d*N*N) - Round(β*d*N*N)** are randomly set of type **G**. The properties of the initial segregate pattern are similar to that in Figure 1b, but the pattern itself is constructed by locating the **B**-agents on the **NxN** grid at 100% density beginning from its left border, column after column, from top to bottom within the column, and by **G**-agents, symmetrically located beginning from the right border, column after column, from the bottom up within the column.

**2.2.   Characterizing model patterns**

2.2.1. Moran **I** as an index of segregation

To characterize the model's spatial pattern we define spatial variable **x$_h$**: **x$_h$ = 1** if **h** is occupied by the **B**-agent and **x$_h$ = 0** otherwise; **x$_h$** is ignored if **h** is empty. We employ Moran index **I** of spatial association to estimate segregation level (Zhang and Linb 2007):

$$I = \frac{N \sum_i \sum_j w_{ij}(x_i - \bar{x})(x_j - \bar{x})}{(\sum_i \sum_j w_{ij}) \sum_i (x_i - \bar{x})^2} \quad (2)$$

Where $x_i$, $x_j$ denote the values of $x_h$ in cells $i$, $j$; $M = \mathbf{Round(d*N*N)}$ is the overall number of occupied cells, $\bar{X}$ is the mean value of $x_h$ over the occupied cells, $w_{ij} = \mathbf{1}$ if $j \in \mathbf{U(i)}$, and $w_{ij} = \mathbf{0}$, otherwise.

We do not account for the bias in (2), which is close to **-1/(N-1)** (Anselin 1995) in order to obtain the values of **I** that exactly repeat those calculated by the popular GeoDa software (https://www.geoda.uiuc.edu/).

Simulation modeling has a basic problem of revealing *all possible* dynamic regimes for a given set of parameters. Easily missed in simulation is stable equilibrium with narrow domain of attraction, especially if there is another stable equilibrium whose domain of attraction is wide. In attempting to avoid mistakes of this kind, we performed extensive numeric experiments with different sets of parameters before conducting investigation of the model. Where possible, we combined this analysis with the analytical reasoning (see Appendix).

Our model is stochastic, and below we call "persistent" the patterns whose characteristics remain within the same interval of variation for a period of time which is many times longer than the time it took the system to converge to this pattern.

2.2.2. The criteria of patterns' randomness and segregation

To recognize random-like patterns we employ a permutation test for the 100x100 or 50x50 city. For the 100x100 city, the 95% confidence interval of Moran **I** is close to (-0.002, 0.002), and 99% confidence interval is close to (-0.004, 0.004), while for the 50x50 city these intervals are about (-0.02, 0.02) and (-0.03, 0.03), respectively. In the following we thus consider the 50x50 pattern non-random in case of Moran **I** > 0.03[3]. Note that Moran **I** can be insufficient to fully characterize the non-random patterns and we would employ, when necessary, additional indices.

## 2.3. The basic view of the Schelling model dynamics

Our numeric experiments demonstrate that for initially random pattern, depending on **F**, the model's dynamics can be considered as comprising of two phases. These phases can be qualitatively characterized by the dynamics of the numbers of happy agents, i.e. those, located in neighborhoods where $f_a(h) \geq F$. For the representative case of an initially random pattern and the **B:G** ratio equal to 1:1, the phases are as follows (Figure 3):

The first – *fast phase* - is characteristic of the pattern with many unhappy agents, as at the start of the simulation. Such patterns always quickly evolve, and during some hundred or fewer iterations the number of unhappy agents drops to almost zero. At the end of the first phase, the pattern remains random-like for low **F**, while for the high **F** the segregate patches appear.

At the second – *slow phase* - the pattern changes slowly. The number of unhappy agents remains close to zero, hence cannot be used for characterizing the pattern. If the pattern exhibits clusters at the end of the first phase, then some of the patches continue to grow during the second phase while the others diminish; the overall number of patches also decreases. The segregate pattern once having emerged is preserved for very long period of time.

---

[3] Formally, for any **F > 0** the pattern is not random; to demonstrate this one needs a "very large" city.

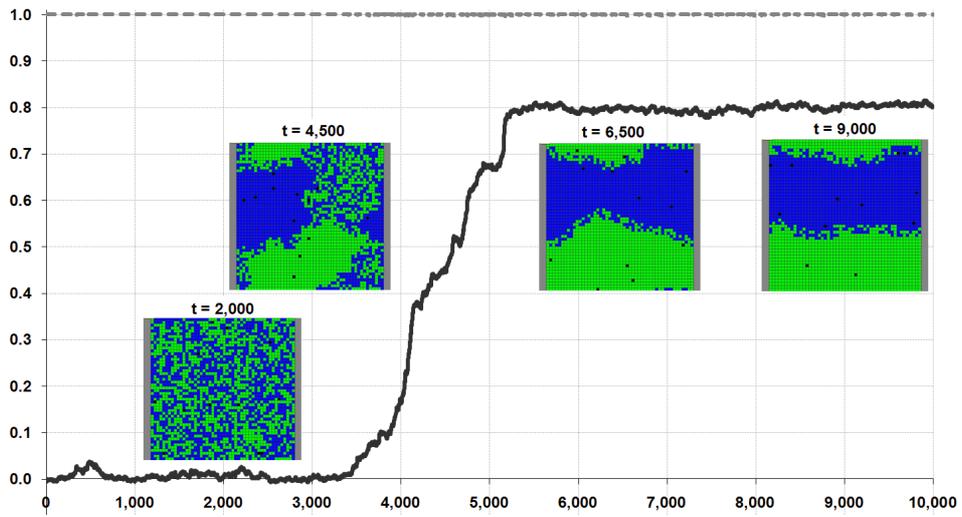

(a)

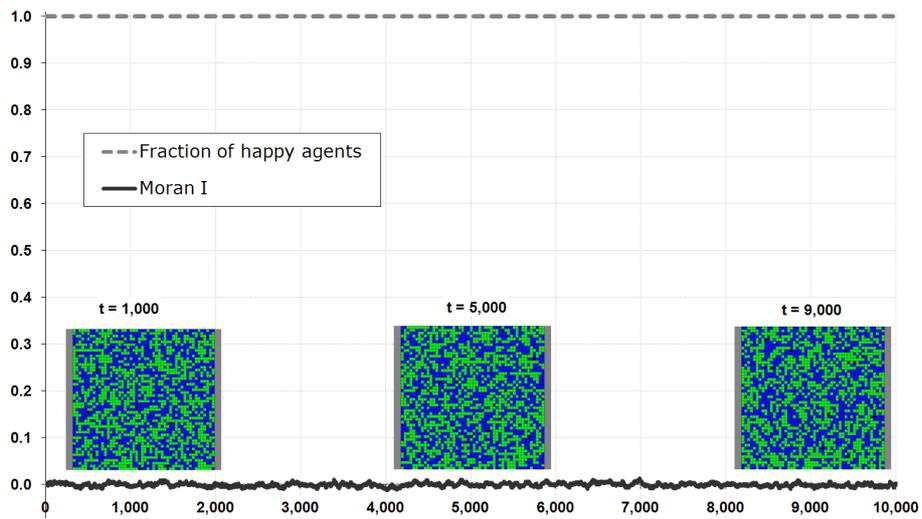

(b)

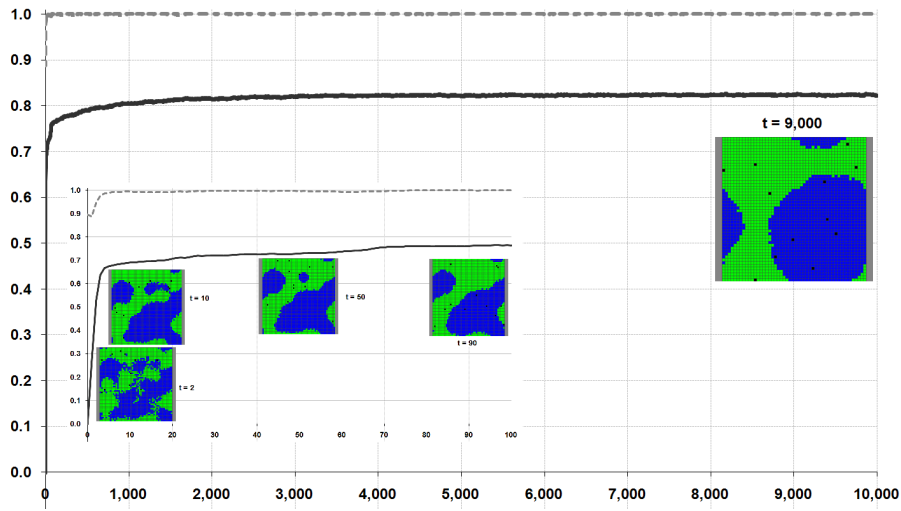

(c)

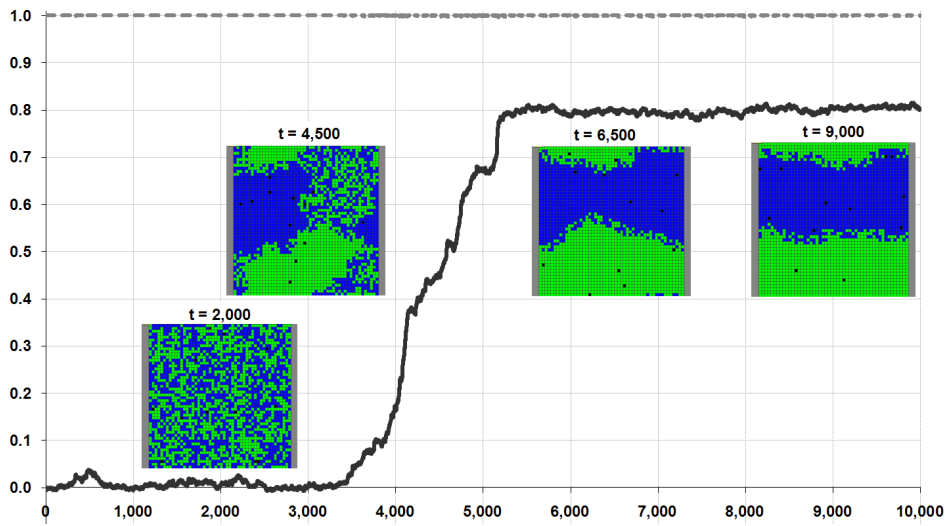

(d)

Figure 3: The model dynamics given by the Moran **I** and fraction of happy agents, and representative patterns for **r** = 3 and characteristic values of **F**: (a) **F** = 8/48 – random pattern; (b) **F** = 20/48 – fast segregation; (c) **F** = 11/48 – boundary value

If the pattern is random-like at the end of the first phase, it remains random unless the values of **F** are the boundary ones that separate between the **F** generating random-like and segregate patterns. For these values of **F** the pattern can undergo a qualitative change, when segregated cluster(s) of agents of one or of both types emerge after very long (and essentially varying in length between the runs) period of randomness (Figure 3c, see (Durrett 1999) for further discussion).

Given **F**, **β**, and **r**, typical duration of the first and second phases for the non-boundary values of **F** is determined by **d** and **m**. For the values of **d** = 0.98 and **m** = 0.01 given in Table 1, the first phase takes at most some hundred iterations. To avoid the problems related to length of this phase we chose the pattern at **t** = 1000 as representing the state of the system at the end of the first phase.

Long-run evolution of the pattern characteristic of the stochastic model should also be treated with care, and typical practice is to run the model for some ten million iterations (Vinkovic and Kirman 2006) in the hope that the pattern at the end of this extremely long period is in a persistent state. In what follows we employ the long runs of **5000/m** time steps that is, of **t** = 500000 for the exploited value **m** = 0.01[4].

### 3. Schelling's basic case - two groups of equal size (β = 0.5)

This case of **β** = 0.5 corresponds to Schelling's original formulation, and our goal is to reproduce the qualitative result, i.e., that for low **F** the residential distribution converges to a random pattern and for high **F** it converges to a segregate pattern. To verify, we performed 30 model runs for each value of **F** given by (1), for **r** = 1, 2, 3 and 4, and for two types of initial conditions, random and segregate. We recorded the pattern and the value $I_{1000}$ of the Moran **I** at the end of the first phase, i.e., at **t** = 1000, and then at **t** = 5000***k**, **k** = 1 to 6. The results are presented in Figure 4 and one can easily note that *the model does reproduce Schelling's dichotomy* - for each **r** and **F** the patterns converges to either random or segregate one.

---

[4] The probability that an agent will not try to migrate for random reasons during 1000 iterations is below $0.99^{1000}$ < 0.00005; for 500000 iterations this probability is practically zero.

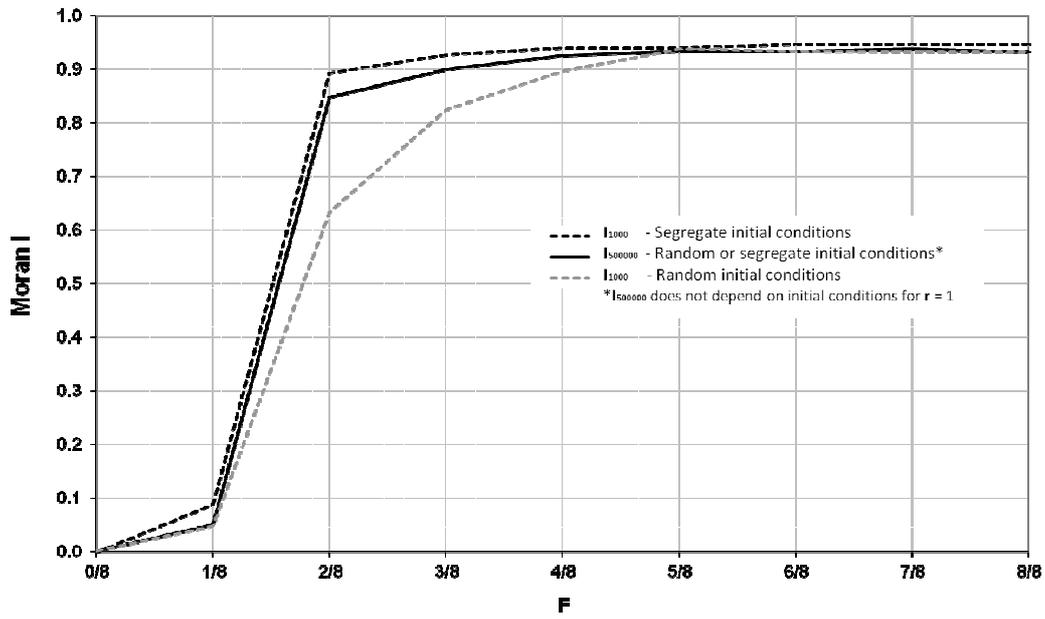

(a)

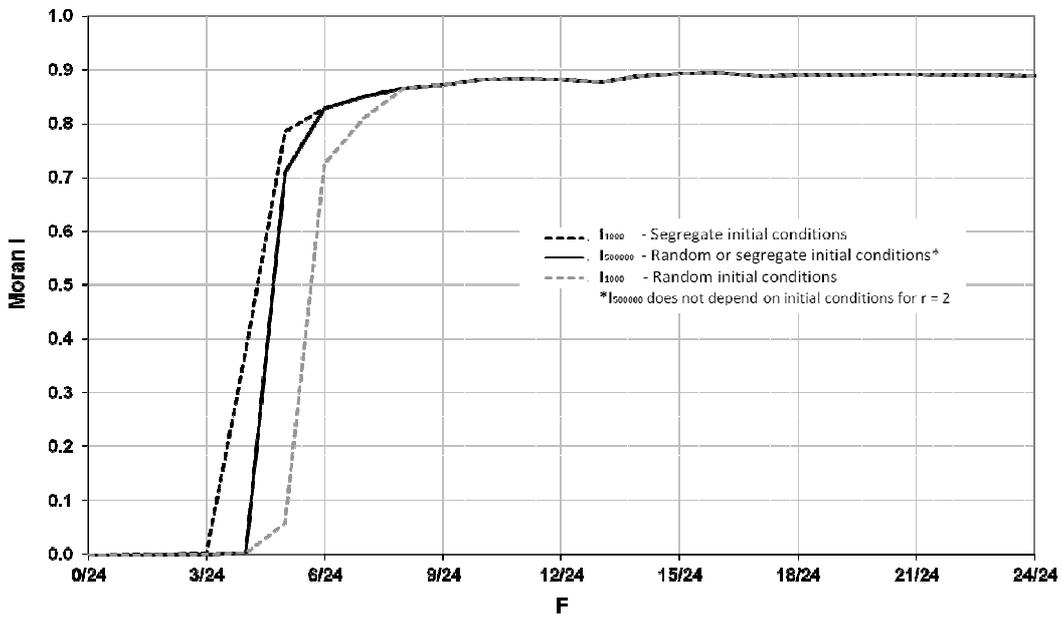

(b)

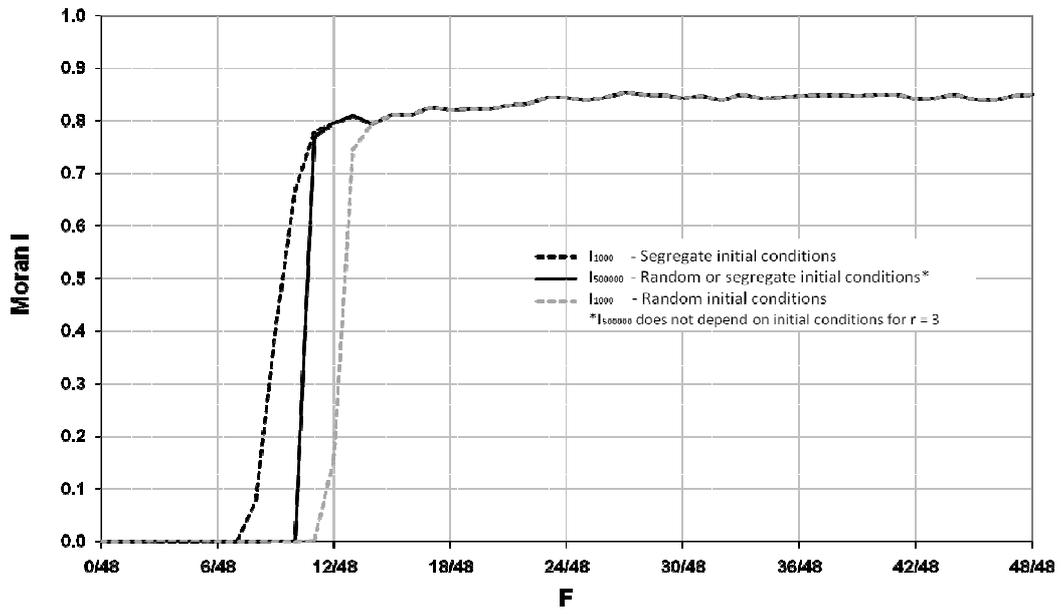

(c)

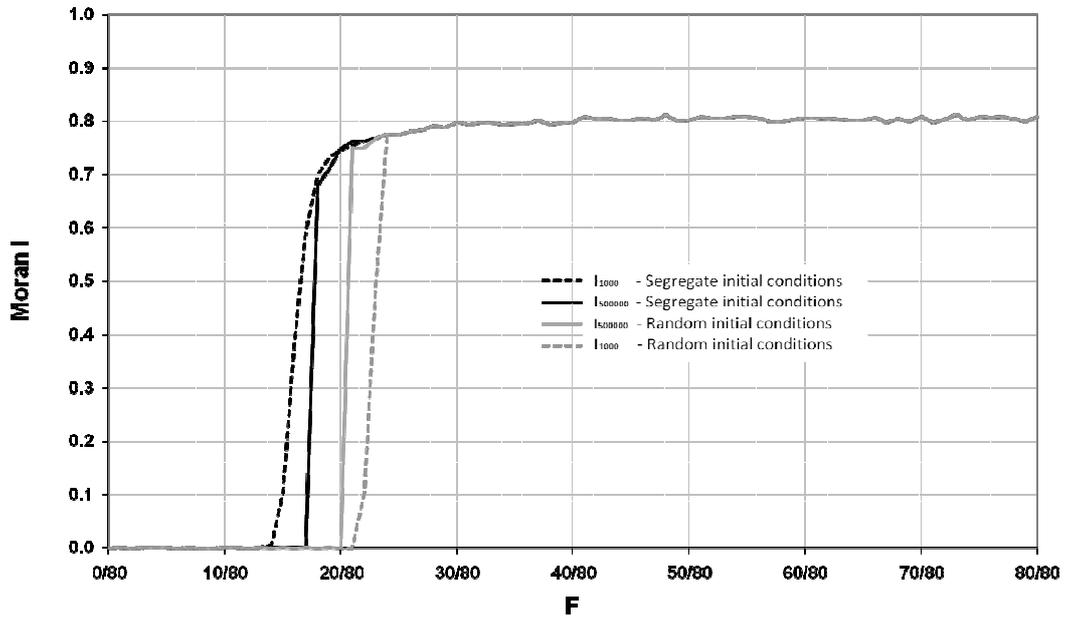

(d)

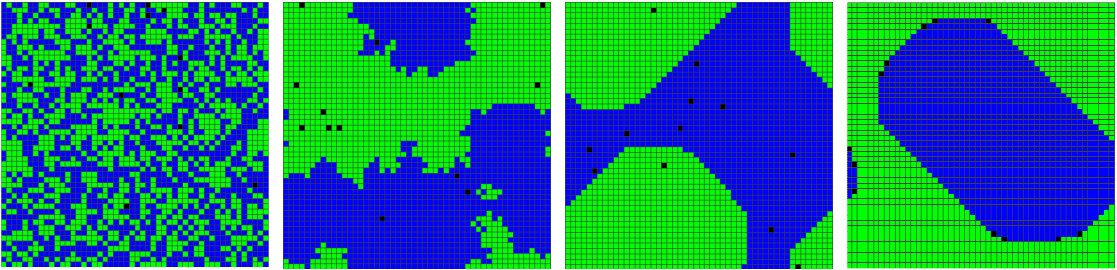

F = 1/8, $I_{500000}$ = 0.001    F = 2/8, $I_{500000}$ = 0.847    F = 4/8, $I_{500000}$ = 0.923    F = 6/8, $I_{500000}$ = 0.928

Figure 4: (a) - (d) **r** = 1 to 4, dependence of the $I_{1000}$ on **F** for random (lower outer curve) and segregate (upper outer curve) initial conditions; dependence of $I_{500000}$ on **F** for random initial (lower inner curve) and segregate (upper inner curve) conditions; (e) **r** = 1, random initial conditions, persistent patterns for **F** = 1/8, 2/8, 4/8 and 6/8, and the corresponding value of $I_{500000}$.

To characterize the model behavior, let us denote as $F_{r,rand,1000}$ the maximal value of **F** that for any initial pattern results in a random-like pattern at **t** = 1000. Respectively, let $F_{r,segr,1000}$ be the minimal value of **F** that for any initial pattern results in a segregate pattern at **t** = 1000. Columns 3 and 4 of the Table 2 presents the values of $F_{r,rand,1000}$ and $F_{r,segr,1000}$ for **r** = 1 to 4.

| Table 2: $F_{r,rand,1000}$, $F_{r,segr,1000}$, $F_{r,rand}$, $F_{r,segr}$ for **r** = 1 to 4. The city is a 50x50 torus | | | | | |
|---|---|---|---|---|---|
| r | n(r) | $F_{r,rand,1000}$ | $F_{r,segr,1000}$ | $F_{r,rand}$ | $F_{r,segr}$ |
| 1 | 8 | 0/8 = 0.0000 | 4/8 = 0.5000 | 1/8 = 0.1250 | 2/8 = 0.2500 |
| 2 | 24 | 4/24 = 0.1667 | 6/24 = 0.2500 | 4/24 = 0.1667 | 5/24 = 0.2083 |
| 3 | 48 | 9/48 = 0.1875 | 13/48 = 0.2708 | 10/48 = 0.2083 | 11/48 = 0.2292 |
| 4 | 80 | 16/80 = 0.2000 | 24/80 = 0.3000 | 17/80 = 0.2125[*] | 21/80 = 0.2625[*] |
| [*]For **F** = 18/80, 19/80, and 20/80, the initially random pattern remains random at t = 500000, while the initially segregated pattern remains segregated at t = 500000. | | | | | |

As may be expected, with the growth of **t** the influence of the initial conditions weakens. Let us denote as $F_{r,rand}$ the maximal **F**-value at which, for any initial distribution of agents, the pattern remains random for **t** → ∞. Similarly, let $F_{r,segr}$ be the minimal **F**-value at which, for any initial distribution of agents, the pattern remains segregate for **t** → ∞. According to our numeric experiments, at **t** = 500000 for **r** = 1, 2, and 3, $F_{r,segr} = F_{r,rand} + 1/n(r)$. That is, in a very long run, the model pattern does not depend on initial conditions for **r** = 1, 2, and 3. For **r** = 4 the situation is more complicated (Table 2). Namely, for **F** = 18/80 to 21/80, the initially random pattern remains random, while the initially segregate pattern remains segregated up to **t** = 500000. On the base of (Durrett, 1999) we expect that 500000 iterations may be yet insufficient for pattern convergence in these cases, but delay the investigation of these dynamics to further studies.

Note that for all **r** the value of $F_{r,segr}$ is lower than **F** ~ 1/3 given by Schelling.

Let us now extend the standard framework and investigate the dynamics of Schelling's model, assuming that the **B**-agents are the minority in the city.

## 4. The minority patterns ($\beta$ < 0.5)

### 4.1. The new – mixed – pattern

The study of the Schelling model dynamics for the case when the B-agents are the minority that is, $\beta$ < 0.5 reveals the major new patterns, which are *neither random nor completely segregate* and have not yet been ascribed to Schelling model. Below we call these patterns *mixed*. In the mixed patterns, part of the area is occupied by the members of majority exclusively, while the rest is occupied by agents of both types (Figure 5a-c).

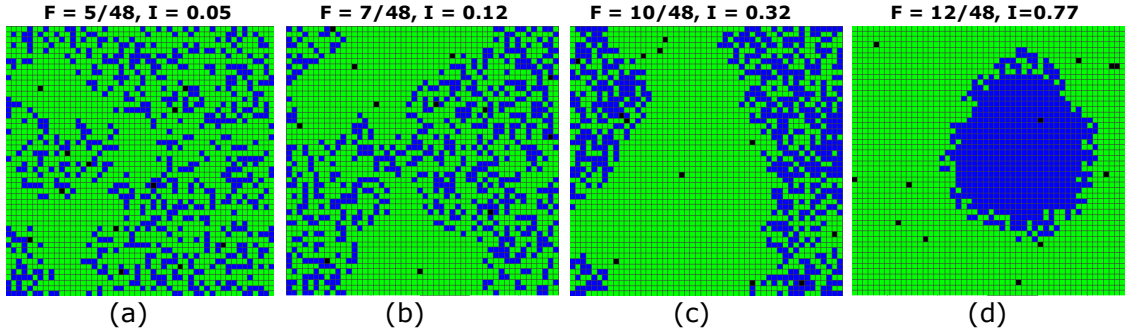

| F = 5/48, I = 0.05 | F = 7/48, I = 0.12 | F = 10/48, I = 0.32 | F = 12/48, I=0.77 |
| (a) | (b) | (c) | (d) |

Figure 5: The persistent patterns for **r** = 3, β = 0.25, and for (a) **F$_k$** = 5/48 (mixed); (b) **F$_k$** = 7/48 (mixed); (c) **F$_k$** = 10/48 (mixed); (d) **F$_k$** = 12/48 (segregate).

### 4.2. Characterization of the mixed patterns

The patterns in Figure 5 clearly show that to investigate the mixed patterns we have to characterize both its segregate and aggregate parts. The Moran **I** calculated for the entire city, as above, is insufficient.

To characterize the mixed pattern let us divide it into four parts (Figure 6). First, let **G^** be the part of the city occupied by the majority group **G** exclusively. We distinguish between internal part **iG^** and the boundary **nG^** of **G^**. A cell **h** ∈ **iG^** if **h** and all cells of the **U(h)** are either occupied by **G**-agents or are empty; cells than belong to neighborhood of some cell from **iG^**, but not to the **iG^** itself, form **nG^**.

Second, let **G⁺** be the complementary to **G^** part of the city. In case of a segregate city, **G⁺** is a homogeneous part occupied by the minority; otherwise **G⁺** represents the heterogeneous part of the city. We define the internal part **iG⁺** of **G⁺** as all cells **h** whose neighborhood **U(h)** does not intersect **G^**, and the boundary **nG⁺** as all cells **h** that belong to the neighborhood of some cell from **iG⁺**, but not to **iG⁺** itself.

Note that the union of **G^** and **G⁺** covers the entire city (Figure 6) and with an increase in **r**, **iG^** and **iG⁺** shrink, while **nG^** and **nG⁺** become wider.

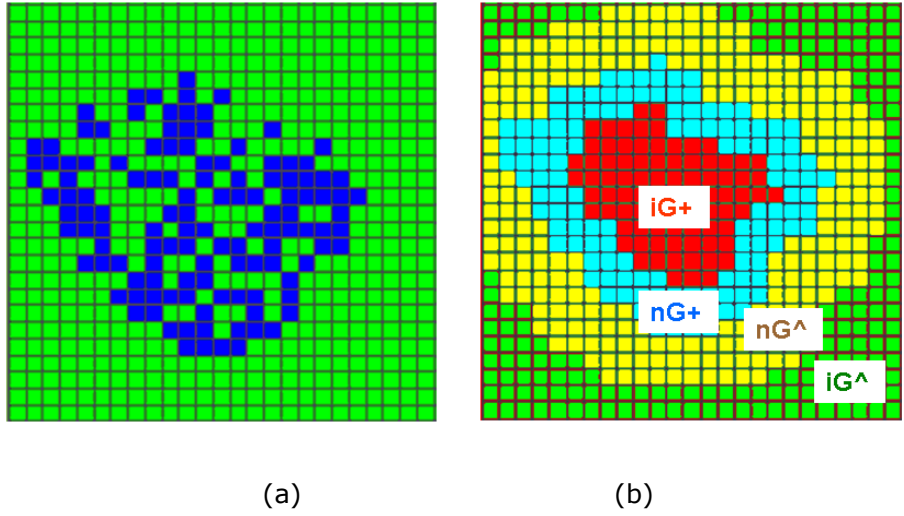

(a)            (b)

Figure 6: The partition of the mixed pattern into four parts (a) pattern (b) **iG^** (green), **nG^** (yellow), **nG⁺**(light blue) and **iG⁺** (red).

In addition to the Moran **I** calculated over the entire city, we characterize the mixed patterns by three measures applied to **G⁺**:

- The value **I⁺** of the Moran **I** calculated over internal part **iG⁺** of **G⁺**.

- The fraction **β⁺** of the minority agents within **G⁺**.

- The area **S(G⁺)** of a mixed part **G⁺**, as a fraction of the entire city area;

### 4.3. Study of the mixed patterns

The hypothesis that can be formulated on the basis of the Figure 5 is as follows. For β < 0.5, three types of persistent patterns - random, mixed and fully segregated - substitute each other with the growth of **F**.

To test this hypothesis let us denote as **F$_{r,rand}$(β)** the maximal **F**-value at which, given β, the persistent pattern is random for any initial distribution of the agents when **t → ∞**. Respectively, let us denote as **F$_{r,segr}$(β)** the minimal **F**-value at which, given β, the pattern remains segregated for any initial distribution of the agents when $t \to \infty$. In what follows we limit ourselves to the cases **r** = 1 to 3.

By definition, for **F** ∈ **[0, F$_{r,rand}$(β)]** the city is random, and for **F** ∈ **[F$_{r,segr}$(β), 1]** the city is segregated. For **r** = 1 to 3, and investigated case of **β = 0.5**, **F$_{r,rand}$(0.5)** and **F$_{r,segr}$(0.5)** are sequential distinguishable values of **F**, i.e., **F$_{r,segr}$(0.5) = F$_{r,rand}$(0.5) - 1/n(r)**. Our hypothesis is that with the decrease of β below 0.5 the interval **[F$_{r,rand}$(β), F$_{r,segr}$(β) − 1/n(r)]** becomes non-zero for any **r**. The width of this interval increases with the decrease of β, and for **F** within this interval the persistent residential pattern is mixed.

To verify this hypothesis we performed series of simulation runs for the 50x50 city. We investigated the values of **r** = 1 to 3, the values of **F** given by (1), and β varying between 0.05 to 0.45, with the step of 0.05. The remaining parameters are as presented in Table 1. We performed regular investigation for the case of **β < 0.5** for random initial conditions, while for the values of **F** close to **F$_{r,segr}$(β)** verified that the persistent pattern for segregate initial pattern is the same as for the random one. For each set of conditions one run was performed and the persistent pattern was estimated in 5000 iterations.

The dependences of the Moran **I** for the entire city, and of three characteristics of the heterogeneous parts – **I$^+$**, **β$^+$**, and **S(G$^+$)** as dependent on **F** and β, for **r** = 3, are presented in Figure 7 (the results for **r** = 1 to 2 are qualitatively similar):

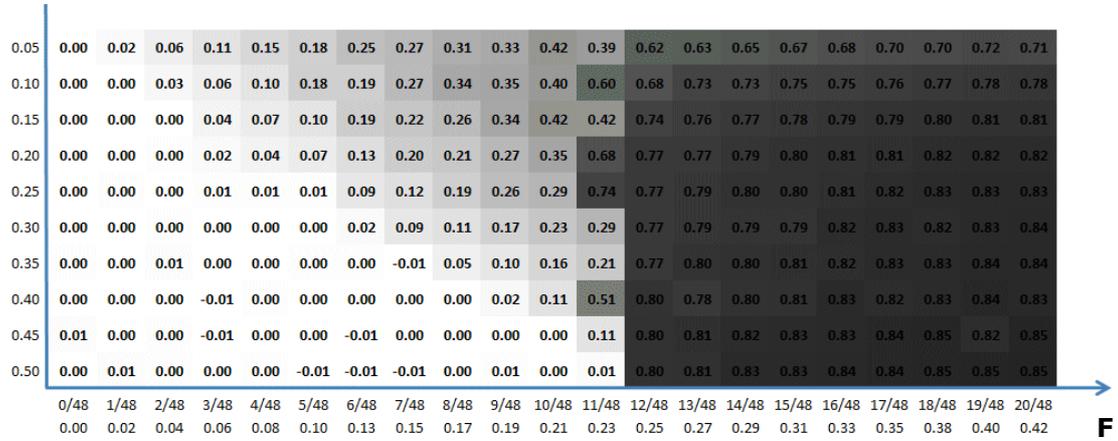

(a)

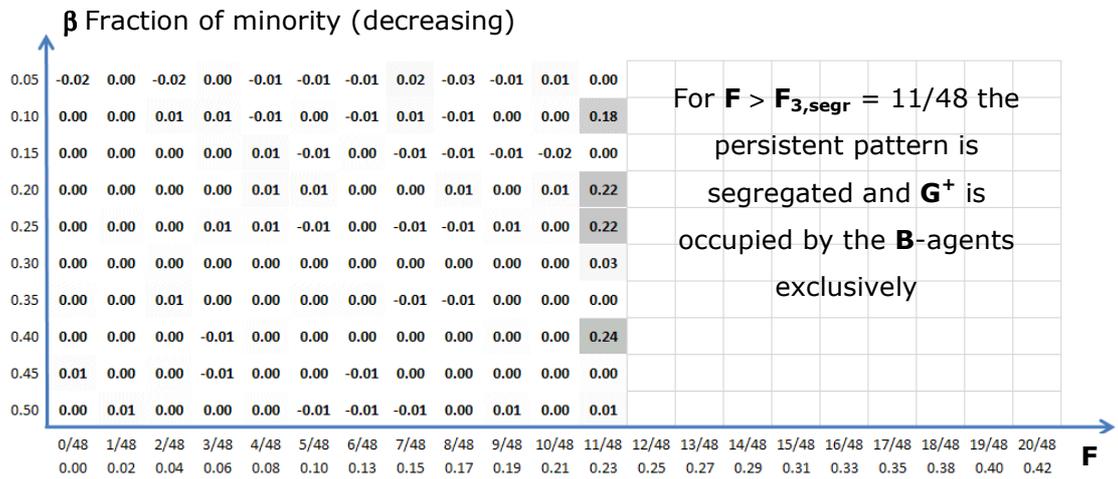

(b)

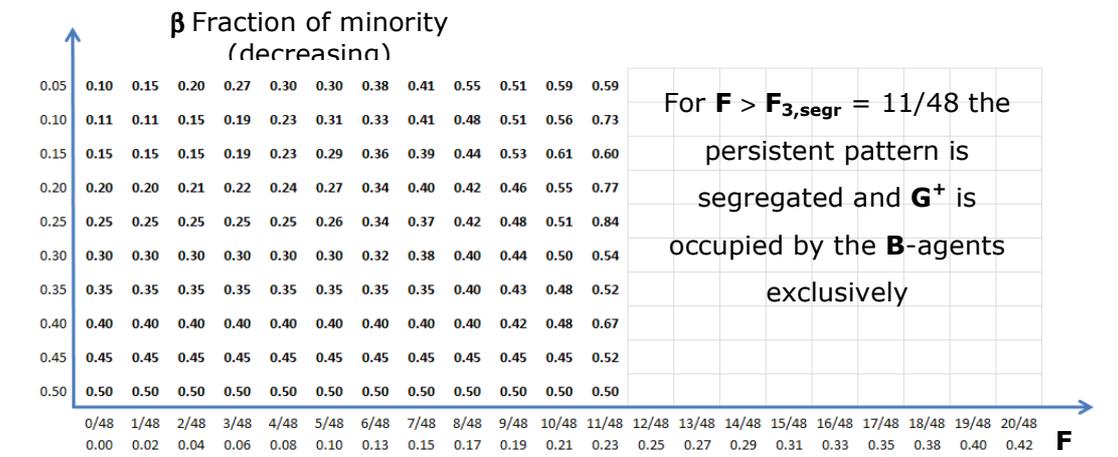

(c)

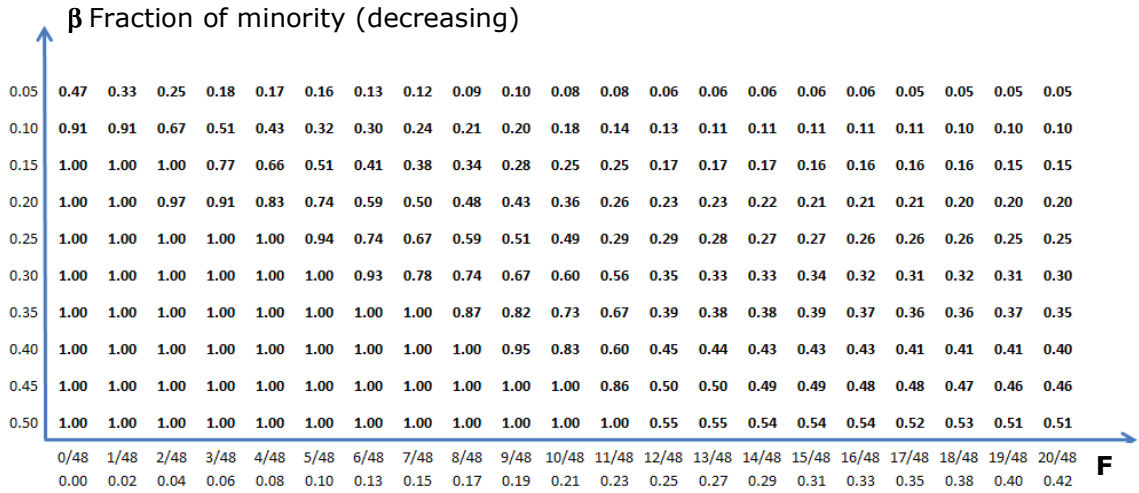

(d)

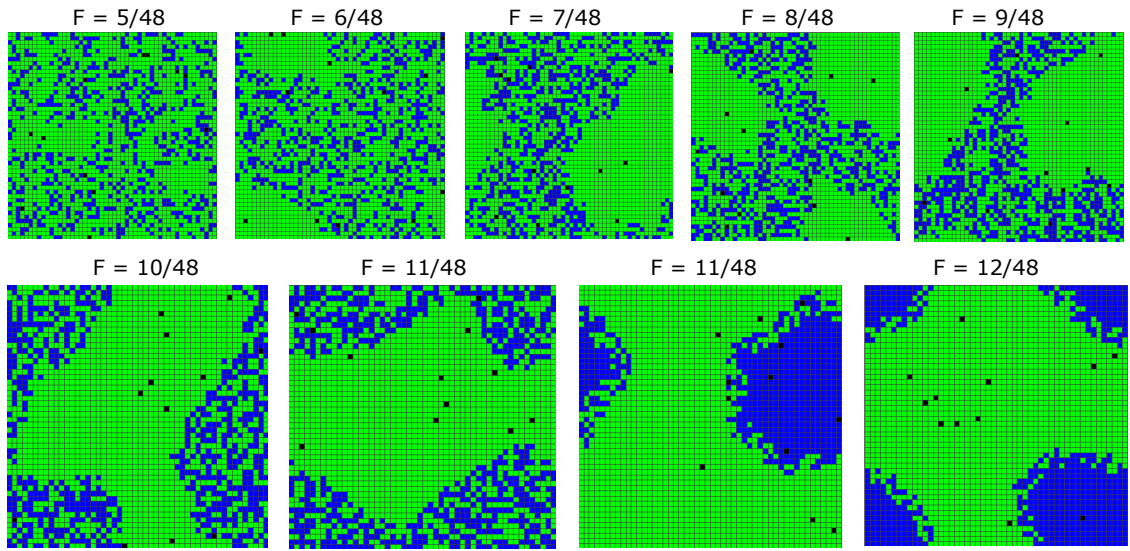

(e)

Figure 7: Characteristics of persistent patterns for **r** = 3 depending on minority fraction **β** and **F**, **β** = 0.05 to 0.5 with step of 0.05, **F** = 0/48 to 20/48 with step of 1/48, and random initial conditions. (a) Moran **I** for the entire city; (b) The value **I⁺** of Moran **I** for the internal part **iG⁺(β)** of the heterogeneous area, red in Fig. 6; (c) The fraction of the **B**-agents within **G⁺(β)**, red and blue in Fig. 6; (d) The area of **G⁺(β)** (red and blue in Fig. 6) as a fraction of the entire city area; (e) persistent patterns for **β** = 0.25, **F** = 5/48 to 12/48 with step of 1/48, enlarged for **F** close to $F_{r,segr}$. Note two qualitatively different patterns obtained for $F = 11/48$, which explain non-zero Moran $I^+$ for $F = 11/48$ and some values of $\beta$.

The analysis of simulation results leads to the following conclusions:

- The values of $F_{r,segr}(\beta)$ do not depend on $\beta$, so $F_{r,segr}(\beta) = F_{r,segr}$ (Figure 7a).

- The lower $\beta$, the wider the interval $[F_{r,rand}(\beta), F_{r,segr} - 1/n(r)]$ ( Figure 7a, gray).

- Agents' spatial pattern within the internal heterogeneous part $iG^+$ is random (Figure 7b).

- The density $\beta^+$ of the **B**-agents in $G^+$ is always higher than $\beta$, grows with the increase in **F** within $[F_{r,rand}(\beta), F_{r,segr} - 1/n(r)]$ and passes $\beta^+ = 0.5$ for $F = F_{r,segr} - 1/n(r)$ (Figure 7c).

- The area of $G^+(\beta)$ as a fraction of the entire city area decreases with the increase in **F** within the interval $[F_{r,rand}(\beta), F_{r,segr} - 1/n(r)]$, , approaching $\beta$ when **F** exceeds $F_{r,segr}$ (Figure 7d, e). Note that in the case $F = 11/48$, 5000 iterations are insufficient for convergence and both mixed and close to segregate $G^+$-patterns are observed.

The results of the analytical study confirm simulation results (see Appendix)

## 5. Discussion and conclusions

Basically, our description of the persistent patterns just extends Schelling's initial observations (Schelling 1978), p. 147-148). Given the neighborhood's radius **r** and the fraction of minority $\beta$, the model's pattern converges to one of three (and not two, as claimed by Schelling) persistent states. The state is defined by **F**, and the **[0, 1]** interval of **F** variability can be divided into three parts: $F \in [0, F_{r,rand}(\beta)]$ entails the random pattern, $F \in (F_{r,rand}(\beta), F_{r,segr})$ the mixed pattern, and $F \in [F_{r,segr}, 1]$ the segregate pattern.

For $\mathbf{F} \in (\mathbf{F_{r,rand}(\beta), F_{r,segr} - 1/n(r)}]$ the pattern consists of a homogeneous part, which is occupied by the majority, and a mixed part, which is randomly occupied by an equal number of agents of both types. The density of the minority in the mixed part is always higher than β; given β, the density of the minority within $\mathbf{G^+}$ grows with the growth of **F**.

For the investigated values of the population density **d** and probability of spontaneous migration **m**, convergence to the persistent pattern takes several hundred iterations for all **F** besides the values that are just below $\mathbf{F_{r,segr}}$. In the latter case the convergence requires more time and, moreover, for **r** = 4, and three consecutive values of $\mathbf{F_k}$ just below $\mathbf{F_{4,segr}}$ = 21/80, the residential pattern remains, at t = 500000, random or segregate depending on whether initial conditions are random or segregate (Table 2).

Let us discuss the meaning and possible interpretation of the results in regard to our starting point: residential distribution in the city.

### 5.1. The robustness and sensitivity of the Schelling model

After being formulated analytically or as a computer program, the model embarks on its own life. And if the source of researcher's inspiration is a "soft" social science, this question arises: does the model, as a formal creature, reflect the general vision of the real phenomenon? The answer is based on *robustness* of model outcomes to the changes in model rules. Unlike sensitivity analysis, which can be employed formally after the model is expressed as a set of equations/rules/lines of computer code, the study of robustness depends on the informal spectrum of "possible changes in the model rules". When studying Schelling model in respect of urban residential dynamics we thus have to rely on intuition.

Let us consider in this respect the consequences of three qualitative assumptions we made:

− Low fraction of empty cells **d**;

− Non-zero probability of spontaneous migration **m**;

− Satisficer behavior of agents

*Low **d** and non-zero **m**:* The values we have employed − **d** = 0.98 and **m** = 0.01, were chosen to reflect the reality, where the fraction of attractive empty apartments (**1** − **d**) is always low, while the reasons for residential migrations, besides reaction to neighbors, are multiple (**m** > 0).

However, the non-zero value of **m** has essential formal consequences. Namely, it prevents the model's pattern from stalling in the "solid" state (Vinkovic and Kirman 2006), when none of the agents can improve their utility. Opposite to the case **m** > 0, for **m** = 0 Schelling model generates the variety of the solid patterns, which in addition depends on **d** (Figure 8). Generally, these solid-state patterns are very sensitive to parameters and to the "arbitrarily" chosen rules, as whether the fraction of friends is calculated by dividing by the size of entire neighborhood, as in Vinkovic and Kirman (2006), or by the number of occupied cells in the neighborhood. This vulnerability of the model's results regarding assumptions that can be never tested impels us to conclude that the rules that make possible solid-state dynamics are irrelevant to a social interpretation of the Schelling model.

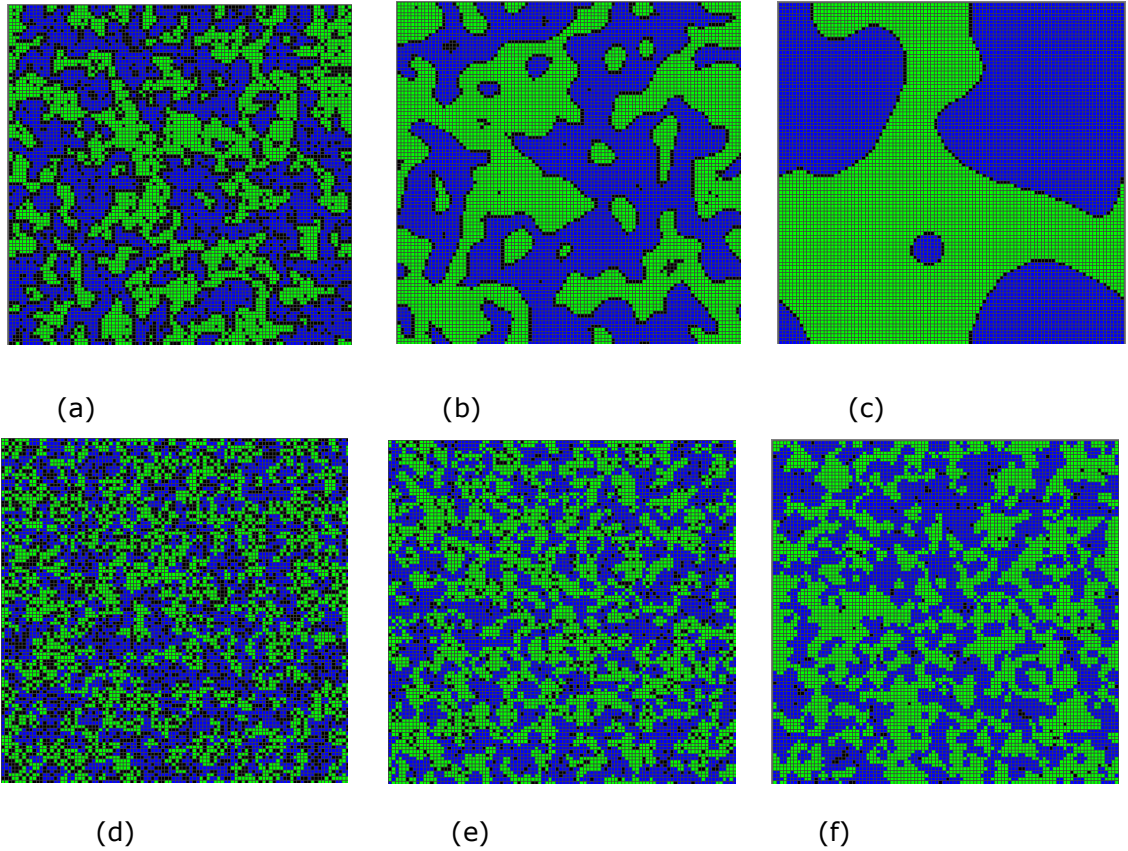

Figure 8: Persistent patterns for **m** = 0. Top: **r** = 1, **F** = 5/8, (a) **d** = 0.7, (b) **d** = 0.9, (c) **d** = 0.98. Bottom: **r** = 1, **F** = 3/8, (d) **d** = 0.7, (e) **d** = 0.9, (f) **d** = 0.98

We tested the sensitivity of the model's persistent patterns to variations of **m** in the interval (0.01, 0.05). Basically the persistent patterns do not change, while the increase in **m** accelerates, and the decrease in **m** decelerates convergence to the persistent pattern. We did not investigate the sensitivity of the model's results to **d**.

*Satisficer behavior of agents*: According to Step 2 of the agent's behavioral rules, the model's agents are satisficers (Simon 1982), that is, they do not distinguish among potential locations provided the number of friends there is above the threshold. To investigate the importance of this assumption, we substituted the satisficer's choice rule with the maximizer's one: the more friends the better. Formally, the maximizer agent **a** estimates the utility of vacancy **h** as **f$_a$(h) − F**. The consequence of this assumption is quite expectable: maximizers always segregate. That is, irrespective of **F**, the pattern always converges to complete segregation.

Based on all these, *we consider close to unit **d**, non-zero **m**, and satisficer behavior of agents as necessary setting for social interpretation of the Schelling model outcomes.*

The lengthy time of convergence to the persistent pattern for the values of **F** close to **F$_{r,segr}$** has essential consequences when the demand for friends in neighborhood **F** *changes over time*. Let us consider the dynamics of the residential pattern in the city, where the preferences of the residents evolve.

## 5.2. Dynamics of residential distribution in the city where residents' tolerance level evolves

Let us consider the city in which initially "good" relationships between the agents of the **B**- and **G**-groups, formally expressed as zero or low **F** deteriorate and the value of **F** grows. To illustrate, let us consider the case of **r** = 3 and assume that initially zero **F** grows at a rate Δ**F**=1/48 (minimal feasible change) every 20 iterations.

With the growth of **F**, the pattern initially does not change at all (Figure 9, lower curve until **F** exceeds **F$_{r,segr}$**. As one can see, for Δ**F**=1/48 per 20 iterations the pattern remains random until **F** reaches 15/48 (i.e., essentially above **F$_{3,segr}$**) and then segregates in 60 iterations only, i.e. externally seems quite sudden.

However, should relations improve again, a long time must pass for the segregated pattern to dissolve. As can be seen from the upper curve in Figure 9, the segregated pattern will persist almost unchanged for the first 200-300 iterations, until the value of **F** will not decrease to very low values, much below **F$_{3,rand}$**. It will take at least 500 iterations to get back to the "integrated" pattern.

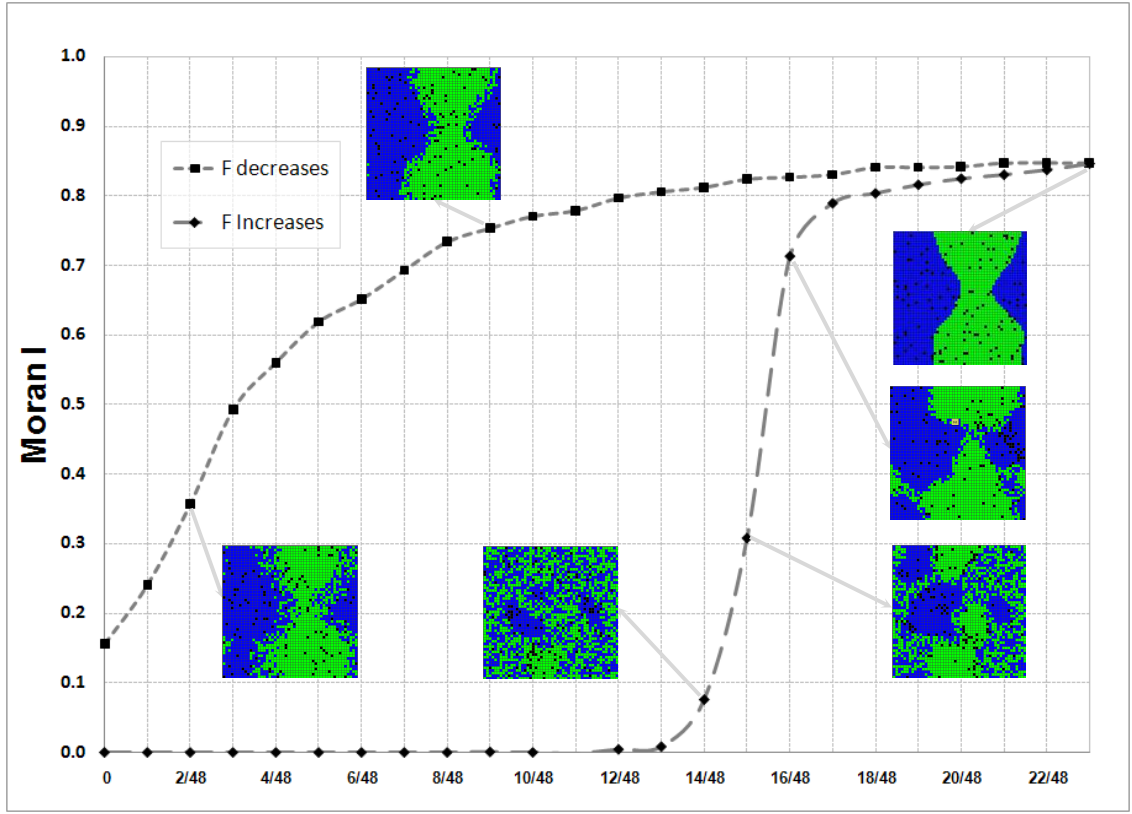

Figure 9: Lower curve – dynamics of the residential pattern for **F** starting at zero and increasing at a rate of 1/48 at every 20$^{th}$ iteration. Upper curve - dynamics of the residential pattern for **F** starting with 23/48 and decreasing 1/48 at every 20$^{th}$ Iteration.

### 5.3. Further studies

The minority-majority case is a first step toward full investigation of the Schelling model dynamics. The next step should be investigation of model's dynamics and persistent patterns in case of different demands of **B**- and **G**-agents regarding the fraction of friends in the neighborhood (Schelling 1978, p. 153-154). In the next paper (in preparation), we consider the case of $F_B \neq F_G$ and more general case of heterogeneous in regard to the **F**, **B**- and **G**-groups.

**Appendix: Mean-field view of the persistent minority-majority patterns**

A rough theoretical view of the minority-majority pattern can be obtained if we ignore spatial dependency between the overlapping neighborhoods. To make the formulae shorter, we assume below that **d** = 1.0, i.e., all cells are occupied.

Let us remind that **G^** denotes the homogeneous part occupied by the majority, and **G⁺** is the rest of the city (Figure 6).

Ignoring the overlap between neighborhoods, the probability $p_k$ that for a given fraction of minority population β, the number of **B**-agents within the neighborhood of radius **r** equals to **k** can be estimated on the basis of binomial distribution

$$p_k = C(n(r), k) * \beta^k * (1 - \beta)^{n(r)-k},$$

where **C(n, x)** is a binomial coefficient, and **n(r)** is the number of cells in the neighborhood.

Consequently, the probability $P_k$ that the number of friends in the neighborhood is below or equal to **k** is

$$P_k = \sum_{x \leq k} C(n(r), x) * \beta^x * (1 - \beta)^{n(r)-x} \quad (A1)$$

Based on (A1), let us estimate the area **S(G⁺)** of **G⁺**, the fraction $\beta^+$ of the minority within **G⁺**, and the values of $F_{r,rand}(\beta)$ and $F_{r,segr}$ and compare them to the experimental estimates given in Figure 7.

A.1 Estimates of $\beta^+$ and **S(G⁺)**

Note that given **F**, the minimal number of friends in a fully occupied neighborhood of a persistent pattern should be **k(F) = F*n(r)**, i.e. every **B**-agent located at **h** within **G⁺** has at least **k(F)** friends in **U(h)**. That is, for a persistent pattern, given **k(F)**, the probability $P_{k(F)-1}$ that there are **k(F) − 1** or fewer friends in **U(h)** is close to zero. Below we set "close to zero" as lower than **1/β*N*N**, i.e., on average less than one **B**-agent in the city resides in a neighborhood with less than **k(F)** friends. Substituting **1/β*N*N** into (A1) we obtain

$$1/\beta*N*N > \sum_{x \leq k(F)-1} C(n(r), x) * (\beta^+)^x * (1 - \beta^+)^{n(r)-x}, \quad (A2)$$

where $\beta^+$ is fraction of minority within **G⁺**.

Given **k(F)**, minimal **β⁺(k(F))** satisfying (A2) provides the lowest possible estimate of the density **β⁺** of **B**-agents in **G⁺**, and the maximal possible area of **G⁺** can be estimated as **β/β⁺(k(F))**. The theoretical estimates in comparison to the estimates obtained in numeric experiments for **r** = 3, **β** = 0.25 and meaningful **F**, are presented in Table A1.

| Table A1. Theoretical and numeric estimates of **β⁺** and **S(G⁺)** for **r** = 3, **β** = 0.25 | | | | |
|---|---|---|---|---|
| **F** | Theoretical estimates | | Numeric estimates | |
| | **β⁺(k(F))** | **S(G⁺) = β/β⁺(k(F))** | **β⁺** | **S(G⁺)** |
| 4/48 | 0.275 | 0.909 | Random pattern | Random pattern |
| 5/48 | 0.305 | 0.820 | 0.26 | 0.94 |
| 6/48 | 0.335 | 0.746 | 0.34 | 0.74 |
| 7/48 | 0.360 | 0.694 | 0.37 | 0.67 |
| 8/48 | 0.385 | 0.649 | 0.42 | 0.59 |
| 9/48 | 0.410 | 0.610 | 0.48 | 0.51 |
| 10/48 | 0.435 | 0.575 | 0.51 | 0.49 |
| 11/48 | 0.460 | 0.543 | Segregate pattern | Segregate pattern |
| 12/48 | 0.480 | 0.521 | Segregate pattern | Segregate pattern |

A.2. Estimates of the **F$_{r,segr}$**

To estimate **F$_{r,segr}$** let us consider the persistent mixed pattern, in which the **B**-minority is entirely located in **G⁺**. At this stage the fraction of the **B**-agents in **G⁺** equals **β⁺(k(F))** or is higher. Let us now note that for values of **β⁺** larger than 0.5,

**G⁺** cannot be heterogeneous, and, thus should consist of the minority agents only. Indeed, in case of heterogeneous **G⁺**, the neighborhood of the **B**-agent located on the boundary of the **G⁺** cannot contain more than half **B**-agents. That is why the value of $F_{r,segr}$ is the value of **F** that entails $\beta^+ = 0.5$.

In case of **r** = 3, the value of $\beta^+(k(F)) = 0.5$ is achieved for the **k(F)**-value after the last given in Table A1, i.e., for **k(F)** = 13, thus providing the theoretical estimate $F_{3,segr}$ = 13/48. This estimate is the next after the numerical estimate of $F_{3,segr}$ = 12/48. Note, that just as in the numeric experiments, the theoretical estimate does not depend on β.

A.3. Estimate of the $F_{r,rand}(\beta)$

Given β, let us calculate maximal **k(F)** that yet preserves close to zero $P_{k(F)}$ in (A2); this **k(F)** can serve as an estimate of $F_{r,rand}(\beta)$. Employing **1/β*N*N** as "close to zero", let us fix β while varying **F**, and solve equation (4) in regard to **k(F)**:

$$1/(\beta*N*N) = \sum_{x \leq k(F)} C(n(r), x) * (\beta)^x * (1-\beta)^{n(r)-x} \quad (A3)$$

The obtained values of the **k(F)** provide estimates of $F_{3,rand}(\beta)$ which are quite close to the numeric ones (Table A2)

Let us note that despite the arbitrary choice of the "less than one cell" condition of **1/(β*N*N)** in (A2) and (A3), the sensitivity of the results to this value is low. The results do not change if we employ a "half-cell", or even a "tenth of a cell" condition.

| Table A2. Theoretical and numeric estimates of $F_{3,rand}(\beta)$ | | |
|---|---|---|
| β | Theoretical estimate | Numeric estimate |
| 0.05 | 0/48 | 0/48 |
| 0.10 | 0/48 | 0/48 |
| 0.15 | 0/48 | 2/48 |
| 0.20 | 1/48 | 2/48 |
| 0.25 | 3/48 | 4/48 |
| 0.30 | 4/48 | 5/48 |
| 0.35 | 6/48 | 7/48 |
| 0.40 | 8/48 | 8/48 |
| 0.45 | 10/48 | 10/48 |
| 0.50 | 10/48 | 10/48 |